\documentclass[%
reprint,
showpacs,
preprintnumbers,
amsmath,amssymb,
prc,
floatfix,
a4paper,
]{revtex4-1}

\usepackage{graphicx}
\usepackage{dcolumn}
\usepackage{bm}

\begin{document}

\preprint{APS/123-QED}

\title{Region of hadron-quark mixed phase in hybrid stars}


\author{Tomoki Endo}
\email{endo@t.kagawa-nct.ac.jp}
\affiliation{Division of Physics, Department of General Education, Kagawa National College of Technology, 355 Chokushi-cho, Takamatsu, Kagawa 761-8058, Japan}


\date{\today}

\begin{abstract}
Hadron--quark mixed phase is expected in a wide region of
 the inner structure of hybrid stars. 
 However, we show that the hadron--quark mixed
 phase should be restricted to a narrower region to because of the charge screening effect.
 The narrow region of the mixed phase seems to explain physical phenomena of
 neutron stars such as the strong magnetic field and glitch phenomena, and it would give a new cooling
 curve for the neutron star.
\end{abstract}

\pacs{Valid PACS appear here}

\maketitle


It is widely believed that there should be quark matter in the inner core
 region of neutron stars because the central density of neutron stars is sufficiently high for
 nucleons to dissolve into more elementary particles, namely quarks. 
 Nowadays we consider that compact stars consist of not only nuclear matter but also other
 matter such as hyperons and quarks. We call such stars {\it hybrid stars}.

 Although deconfinement phase transition is still not clearly understood, many
 authors have studied the transition 
 through modeling and first-principle calculations such as those of lattice QCD.
 Properties of quark matter have been 
 actively studied theoretically in terms of quark--gluon
 plasma, color superconductivity \cite{alf1,alf3} and magnetism
 \cite{tat1,tat2,tat3},
 and experimentally in terms of relativistic
 heavy-ion collisions \cite{rhic} and early-universe and compact stars
 \cite{mad3,chen}. 
 Such studies are continuing to provide exciting results \cite{risch}.

Because many theoretical calculations have suggested that deconfinement 
transition is of first order
at low temperature and high density \cite{pisa,latt},  
 we assume it to be a first-order phase transition here.
 The Gibbs condition (GC) \cite{gle1}
 then gives rise to various structured mixed phases (SMPs).
 The SMPs suggested by Heiselberg et al.\ \cite{pet} and Glenndening
 and Pei \cite{gle2} suggest a crystalline structure of the mixed phase
 in the core region of hybrid stars.  Such structures are
 called ``droplets'', ``rods'', ``slabs'', ``tubes'', and ``bubbles''. 
 These SMPs exist within a wide density range 
if we take a moderate value of the surface tension.
However, if the surface effect is strong, the SMPs are limited to a narrow density region. 
On the other hand, Voskresensky et al.\ \cite{vosk} reported the charge screening
 effect for several cases of droplets and slabs. They showed that
 even if we take a moderate value of the surface tension, SMPs cannot exist
 because of the charge screening effect.
They referred to this phenomenon as {\it mechanical instability}.
However, they used a linear approximation to solve the Poisson equation.

 We have presented the equation of state (EOS) for the mixed phase taking
 into account the charge screening effect \cite{end2} without recourse to
 any approximation. 
 The EOS is similar to that obtained from the Maxwell construction (MC).
 The allowed region of the mixed phase should
 then be narrow due to the charge screening effect. 
 
Recently ``black stars" have been proposed by Barcelo et al.\ \cite{barc}, who suggested that quantum effects prevent a star from collapsing into a ``Black hall". 
Therefore, the EOS of hadron matter in the high-density states plays an important role in determining the quantum effects within compact stars. 

In this paper, we apply our EOS to study the structure of 
 hybrid stars and demonstrate how charge screening affects 
 the physical properties of the hybrid stars.


We use the EOS given in our previous paper \cite{end2}, which also presented our framework. 
Therefore, our approach is only briefly explained here.
Thermal equilibrium is implicitly achieved at $T=0$. We consider that
both hadron and quark matter and the mixed phase are $\beta$ stable.
We employ density functional theory (DFT) under the local
density approximation \cite{parr,drez}. 
 
 We consider the geometrically structured mixed phase (SMP) in which one
phase is embedded in the other phase with a certain geometrical form.
We divide the whole space into equivalent Wigner--Seitz cells with 
radius $R_{\mathrm{W}}$. Such cells include an embedded phase with
size $R$. We impose total charge neutrality and chemical
equilibrium to satisfy the GC. Therefore, adjacent
cells do not interact with each other.

The quark phase consists of $u$, $d$ and, $s$ quarks and
electrons in $\beta$ equilibrium. We consider that $u$ and $d$ quarks are massless
and $s$ quarks are massive ($m_s=150$ MeV), and consider one-gluon exchange for
quark interaction.
The hadron phase consists of nucleons and electrons in
 $\beta$ equilibrium. We use
an effective potential parameterized to reproduce the saturation 
property of nuclear matter.
We use empirical values for the binding energy 
($\epsilon_\mathrm{bind} = -15.6$ MeV), saturation density
($\rho_0 = 0.16$ fm$^{-3}$), symmetry energy ($S_0 = 18$ MeV), and compression
modulus ($K_0 = 285$ MeV).

To account for the confinement,
we introduce a sharp boundary between the two phases employing the
bag model with surface tension parameter $\sigma$.
Determination of the surface tension between hadron and quark matter is
a difficult problem;
thus, many authors \cite{pet,gle2,alf2} have treated
the surface tension as a free parameter and observed its effect. We take the same approach in this study.
We use the bag constant of $B=120 \hspace{3pt} \mathrm{MeV/fm^3}$,
which is to the same as that used in Ref. \cite{pet,end2}. 

As we need to account for the Coulomb interaction, the Coulomb
potential appears in the total thermodynamic potential. 

Differentiating the thermodynamic potential $\Omega_\mathrm{total}$ with respect to each constituent
density $\rho_i \hspace{3pt} (i=u,d,s,n,p,e)$ or the Coulomb potential
$V(\vec{r})$ with $r$ being the radial coordinate, we obtain the equations of motion
(EOMs) for $\rho_i$ and $V(\vec{r})$. We then numerically solve these EOMs
under the conditions of chemical equilibrium in each phase and at the hadron--quark boundary.
In particular, we fully solve the
Poisson equation without any approximation. Note that the Poisson equation
becomes highly nonlinear because charged particle densities are complicated functions
of the Coulomb potential.

 Here, we must determine eight variables; i.e., 
the six chemical potentials $\mu_u$, $\mu_d$,
 $\mu_s$, $\mu_p$, $\mu_n$ and
$\mu_e$ and the radii $R$ and $R_\mathrm{W}$.
First, we fix $R$ and $R_\mathrm{W}$.
 We then have four conditions due to the
 $\beta$ equilibrium, expressed by the chemical equilibrium.
Therefore, once the two chemical potentials $\mu_\mathrm{B}$ and
$\mu_e$ are given, 
we can determine the other four chemical potentials 
$\mu_u$, $\mu_d$, $\mu_s$ and $\mu_p$. 
Next, we determine $\mu_e$ according to the global charge neutrality
condition: 
\begin{equation}
 f_V \rho_{\mathrm{ch}}^{\mathrm{Q}} + (1-f_V) \rho_{\mathrm{ch}}^{\mathrm{H}} = 0,
\label{globalneutral}
\end{equation}
where the superscripts ``Q'' and ``H'' denote the quark and hadron phases,
 respectively. The volume fraction is given by $f_V=\left(\frac{R}{R_\mathrm{W}}\right)^d$, and $d$
denotes the dimensionality of each geometrical structure.
At this point, $f_V$ is still fixed.

The pressure obtained from the surface tension is given by 
\begin{equation}
P_{\sigma}= \sigma \frac{d S}{d V_{\rm Q}},
\end{equation}
where $S$ is the surface area of the interface and $V_{\rm Q}$ is the volume of the quark matter. 
We then find the optimal value of $R$ ($R_\mathrm{W}$ is fixed and 
$f_V$ thus changes with $R$) 
using one of the GC :
\begin{equation}
P^\mathrm{Q} = P^\mathrm{H} + P_\sigma.
\label{pbalance}
\end{equation}
The pressure in each phase, $P^\mathrm{Q}$ or $P^\mathrm{H}$, 
is given by the thermodynamic relation 
$P^\mathrm{Q(H)}=-\Omega_\mathrm{Q(H)}/V_\mathrm{Q(H)}$,
where $\Omega_\mathrm{Q(H)}$ is the thermodynamic potential for each
phase.
Finally, we determine $R_\mathrm{W}$ by minimizing the thermodynamic potential. 
Therefore, once
$\mu_\mathrm{B}$ is given, all other $\mu_i$ ($i=u,d,s,p,e$)
can be obtained along with $R$ and $R_\mathrm{W}$.
Note that we keep the GC  throughout the numerical procedure. 

To elucidate the charge screening effect, we also make the calculation without the
screening effect. We refer to this calculation as the ``no-screening'' calculation and the
calculation with the screening effect as the ``screening'' calculation for convenience.

We then apply the EOS derived in our recent paper \cite{end2} to
the Tolemann--Oppenheimer--Volkov (TOV) equation. 


We present the results for the screening and no-screening cases.
\begin{figure}[htb]
\begin{center}
\includegraphics[width=80mm]{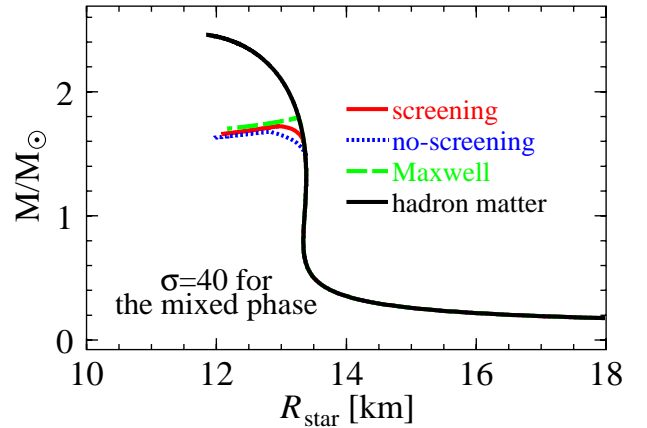}
\caption{ (Color online) Mass--radius relation of stars. The maximum masses
 in the cases of
 pure hadron matter and screened and not screened mixed phases are 2.51, 1.73 and
 1.68 M$_\odot$, respectively. Note that we use a simple model for each
 type of matter. The difference between screened and not screened mixed phases is
 clearly small.}
\label{m-r40}
\end{center}
\end{figure}
Figure\ \ref{m-r40} shows the mass--radius relations of stars in the cases of screening and 
 no screening using the MC and in the case of pure hadron matter.
There is no notable difference in the low-mass region because there is no phase
transition. On the other hand, there is a slight
difference in the region around the maximum mass.
The difference in the maximum masses between the cases of 
screening and no screening 
is about $0.05M_\odot$. 
This difference is small compared with the total mass
of the star. Thus, charge screening 
does not greatly affect the bulk properties of the star.

\begin{figure}[htb]
\includegraphics[width=80mm]{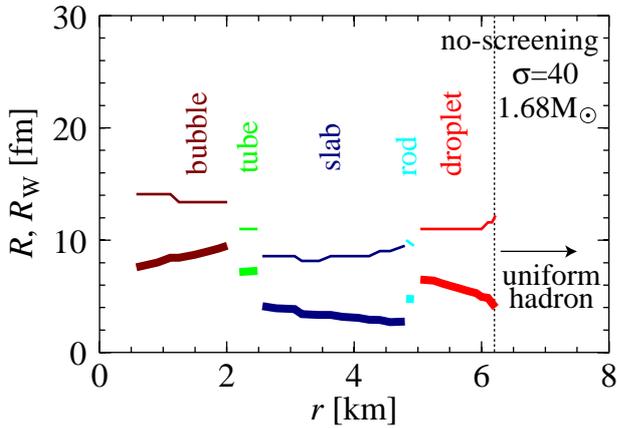}
\caption{ (Color online) Structure size and cell size in the core region of hybrid
 stars in the case of no screening. There is about 6 km of mixed phase in this figure. The radius
 is 12.6 km. Thick lines denote $R$ and thin lines $R_\mathrm{W}$.}
\label{r-ds40no}
\end{figure}
\begin{figure}[htb]
\includegraphics[width=80mm]{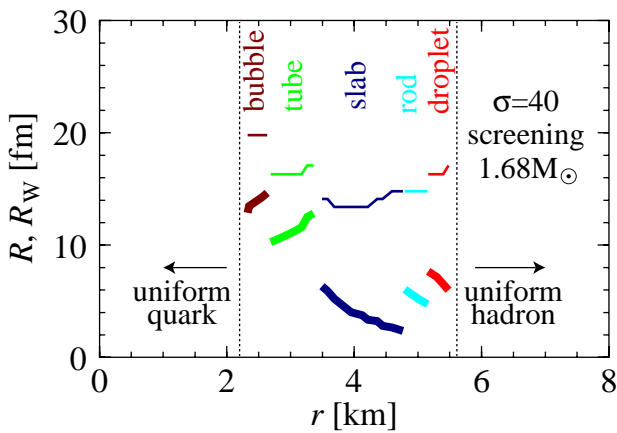}
\caption{ (Color online) Structure size and cell size in the core region of hybrid
 stars in the case of screening. The mixed phase is restricted by the charge screening. The radius is 12.9 km.}
\label{r-ds40sc}
\end{figure}

On the contrary, we find that the inner structure is greatly
affected by charge screening.
Figure\ \ref{r-ds40no} shows the inner core region of a hybrid star in the case without
screening. The mixed phase appears in a wide region of the star. In this case, the mixed phase is spread over a radius of about 6 km, 
and we scarcely see the quark matter phase. 
In the case of
screening shown in Fig.\ \ref{r-ds40sc}, on the other hand,  
there is mixed phase over only 3 km. 
The region of mixed phase clearly
becomes more narrow and there is a quark matter phase in the central region because of
the charge screening.

These results have possible implications for star phenomena.
It has been stated that the strong magnetism of a neutron star is due to 
the spin polarization of quark matter \cite{tat1}, in which case pure quark matter should 
extend far into the core region. 
 Our results suggest
that quark matter could exist in the inner region of compact stars, and it is thus plausible
to attribute the magnetism of compact stars to
spin polarization.

Moreover, Bejger et al.\ \cite{bejg} estimated the
region of mixed phase in a hybrid star and 
suggested that the glitch
phenomenon is caused by the presence of the mixed phase. 
Their results show that the mixed phase is constrained to a narrow region. 
In another studies, gravitational waves have been commonly used to probe density discontinuities \cite{mini}. 
There should be a discontinuity in the hadron--quark phase transition 
if we use the MC. Although the MC is not valid strictly speaking,
the EOS of the mixed phase is close to that obtained using the MC due to the
finite size.
In addition, many theoretical studies have used other models for
the glitch phenomenon and gravitational waves. We cannot simply apply our EOS to
studying these phenomena. However, it is 
interesting to compare our results with those of other studies \cite{kurk}. 

It is, however, plausible to use our results in investigating the cooling problems of neutron stars.  If there is nonuniformity 
in the core region of a hybrid star, it is interesting to investigate whether neutrinos would be scattered by the lumps. 
Cooling would then be prevented and the cooling curve
should change. Although the results strongly depend on the region of
the mixed phase, we suggest new cooling
curves by taking into account the finite-size effects.


In this study, we demonstrated how charge screening 
affects the hadron--quark mixed phase in the core region of hybrid stars.
We examined the effect 
by numerically solving the EOM for particle densities 
and the Poisson equation for the Coulomb potential. 
We demonstrated, by taking various SMPs 
as examples, that the charge screening effect and 
the rearrangement of charge densities play
an important role in determining the inner
structures of hybrid stars.
We applied our EOS to the TOV equation
and found that charge screening did not greatly affect the bulk properties
of the star, such as the radius and mass.
However, we found that the inner structures are greatly affected.
The region of the mixed phase in the star 
is highly restricted by the charge screening.
If we ignore the charge screening effect, the mixed phase would appear in
a wide region. 
On the basis of this result, some authors have suggested 
structured mixed phase with various geometrical structures, simply by including 
the finite-size effects and the surface and Coulomb energies \cite{pet,gle2,alf2,gle3}. 
However, once we take into account the charge
screening effect, the mixed phase is restricted to a narrow region.
In particular, the core could consist of quark matter could appear due to
the charge screening effect.
As another case in which there is more than one chemical potential, kaon condensation  
has been studied \cite{maru} and the results are similar to those in
our paper \cite{end2}. 

 We considered surface tension, but its definite value is not yet clear
 and many authors treat it as a free parameter. 
If we use smaller/larger value of the surface tension, the mixed phase is favored/disfavored.
Here we use the value of 40 $\mathrm{MeV/fm^2}$, but the mixed phase is disfavoed if we use $\sim$90 $\mathrm{MeV/fm^2}$ \cite{pet}.
There are also many
estimations of the surface tension at the hadron--quark
interface in lattice QCD \cite{kaja,huan}, in shell-model
calculations \cite{mad1,mad2,berg}, in model
calculations based on the dual Ginzburg--Landau theory \cite{mond} and in a sigma model \cite{palh}.
If we have a realistic value of the surface tension, we can reasonably deduce SMPs in
the hadron--quark matter phase transition. 
Moreover, we can determine the region of
mixed phase in hybrid stars. We are then able to provide important
information for the EOS of the deconfinement phase transition, 
and clarify the phenomena of neutron stars and ``black stars". 

We have assumed in relation to phenomenological implications that
the temperature is zero. It would be interesting to include the finite-temperature effect. 
We could then study the structure of the proto-neutron star, and present
new cooling curves of the neutron stars.

 In this study, we used a simple model for quark matter to
determine the finite-size effects acting on the SMP in a
hybrid star. 
However,
it has been suggested that color superconductivity is a ground state of quark
matter \cite{alf1,alf2}. To obtain a more realistic picture of
the hadron--quark phase transition, we need to
take into account color superconductivity. 
The description of nuclear matter needs to be improved; for example, we need to take into
account the relativistic mean field theory \cite{shen}. We will then be able to provide 
more realistic results. 


Dr. T.~Maruyama, Dr. S.~Chiba and Dr. T.~Tatsumi are acknowledged for our early work.
This work was supported in part by Division of Physics, Department of General Education, Kagawa National College of Technology.

\end{document}